\documentstyle[11pt,newpasp_piotto,twoside]{article} 
\def\edcomment#1{\iffalse\marginpar{\raggedright\sl#1\/}\else\relax\fi} 
\reversemarginpar 

\begin{document} 
\title{On the Oosterhoff Classification of the Unusual, Metal-Rich 
Globular Clusters NGC~6388 and NGC~6441}

\author{M. Catelan} 
\affil{Pontificia Universidad Cat\'olica de Chile, 
       Santiago, Chile} 
\author{A. V. Sweigart} 
\affil{NASA Goddard Space Flight Center, Greenbelt, MD, USA} 
\author{B. J. Pritzl} 
\affil{National Optical Astronomy Observatories, Tucson, AZ, USA} 
\author{H. A. Smith} 
\affil{Michigan State University, 
       East Lansing, MI, USA} 

\begin{abstract} 
We discuss the Oosterhoff classification of the unusual, metal-rich 
globular clusters NGC~6388 and NGC~6441, on the basis of new 
evolutionary models computed for a range of metallicities. Our 
results confirm the difficulty in unambiguously classifying these 
clusters into either Oosterhoff group, and also question the view 
that RR Lyrae stars (RRL) in Oosterhoff type II (OoII) globular clusters 
can all be evolved from a position on the blue zero-age horizontal 
branch (ZAHB).
\end{abstract}

\section{Introduction} 
NGC~6388 and NGC~6441 are unusual in several respects, 
including the following. (i)~In spite of being metal-rich 
($\rm [Fe/H] \simeq -0.55$~dex), they contain prominent 
extensions to their red HBs---including 
both RRL and extended blue tails (Rich et al. 1997). 
(ii)~As pointed out by Sweigart \& Catelan (1998), the normally 
``horizontal'' part of these clusters' HBs is actually strongly 
tilted, with a $\Delta V \sim 0.5$~mag from the lower part of the 
red HB to the tip of the blue HB---which cannot be accounted for 
by canonical stellar evolution models. (iii)~The mean periods of 
the RRL in both clusters are extremely long for their 
metallicity---longer, in fact, even than in metal-poor, OoII 
globular clusters, thus apparently breaking down the 
traditional Oosterhoff classification scheme and posing yet 
another serious challenge to the models (Pritzl et al. 2000, 
2001, 2002).  

\section{On the Oosterhoff Classification of NGC~6388 and NGC~6441} 
Such peculiarities notwithstanding, it has recently been suggested 
(Clement et al. 2001), on the basis of the RRL's position on the 
$\log\,P-A_V$ plane---i.e., reportedly coincident with the ``line'' 
occupied by ``normal'' RRL stars in OoII globular clusters 
(Clement \& Rowe 2000)---that these globulars should 
be classified as OoII, with the RRL evolved away from a position 
on the blue ZAHB. This suggestion raises some fundamental questions, 
including the following: 

(i)~Do blue HB stars spend enough time within the 
instability strip as they evolve redward to the asymptotic giant 
branch (AGB) to produce the observed number of RRL in OoII globular 
clusters? 

(ii)~Does the predicted 
period-$T_{\rm eff}$ relation for a given OoII cluster depend on 
the stellar distribution along the blue HB? 

(iii)~Is the period-$T_{\rm eff}$ relation independent of metallicity? 

To answer these questions, we have computed new evolutionary tracks 
and HB simulations for OoII globular clusters. Our main results (see 
Pritzl et al. 2002 for a more thorough discussion) can be summarized 
as follows: 

(i)~According to the models, blue HB stars {\em do not} spend enough 
time within the instability strip as they evolve redward to the AGB 
to produce the observed number of RRL in OoII globulars. This confirms 
the well-known arguments of Renzini \& Fusi Pecci (1988) and Rood \& 
Crocker (1989), which unfortunately have been almost entirely 
neglected in the literature since the mid-90's. 

(ii)~The predicted period-$T_{\rm eff}$ relation for a given cluster 
depends somewhat on the stellar distribution along the blue HB, 
leading to the presence of intrinsic scatter in the period-$T_{\rm eff}$ 
plane and to longer mean period shifts for OoII clusters with bluer HBs. 
This has to be taken into account, when attempting to classify 
individual stars into an Oosterhoff type. On the other hand, only 
stars within a relatively narrow range of ZAHB colors, $(B-V)_0 > 0$, 
have a significant chance of becoming redward-evolving RRL. 

(iii)~The predicted period-$T_{\rm eff}$ relation for redward-evolving
RRL is a weak function of metallicity for OoII metallicities, but not 
for the metallicities of NGC~6388 and NGC~6441. 

\acknowledgments This work has been supported by the National Science 
Foundation under grants AST 9528080 and AST 9986943. Support for M.C. 
was provided by Proyecto de Inicio DIPUC~2002-04E. A.V.S. 
acknowledges support from NASA Astrophysics Theory Program proposal 
NRA-99-01-ATP-039.

\end{document}